\documentstyle[12pt]{article}
\input epsf

\begin{document}
\hyphenation{Schwarz-schild}

\thispagestyle{empty}
\rightline{NSF-ITP-94-17}
\rightline{SU-ITP-94-4}
\rightline{hep-th/9402125}

\begin{center}
{\large\bf  STRING THERMALIZATION \break
AT A BLACK HOLE HORIZON} \break

\vskip 0.9 cm
{\bf Arthur Mezhlumian} \footnote{On leave from Landau Institute
for Theoretical Physics, Moscow, Russia.  \hfil\break
\indent\indent Electronic address: arthur@physics.stanford.edu},
{\bf Amanda Peet}
\footnote{Electronic address: peet@dormouse.stanford.edu},

Physics Department \break
Stanford University  \break
Stanford, CA 94305-4060  \break

and {\bf L{\'a}rus Thorlacius} \footnote{Electronic address:
larus@nsfitp.itp.ucsb.edu}

Institute for Theoretical Physics  \break
University of California  \break
Santa Barbara, CA 93106  \break

\end{center}
\vskip0.9cm

\begin{quote}
Susskind has recently shown that a
relativistic string approaching the event horizon of a black hole
spreads in both the transverse and longitudinal directions in the
reference frame of an outside observer.  The transverse spreading can
be described as a branching diffusion of wee string bits.  This
stochastic process provides a mechanism for thermalizing the quantum
state of the string as it spreads across the stretched horizon.
\end{quote}
\vskip1cm

\normalsize

\newpage

\section{Introduction}

Hawking's information paradox \cite{Hawking},  {\it i.e.} the
question whether black hole evaporation can be reconciled with
the unitary evolution of states in quantum theories, has received
considerable attention in recent years.
An attractive resolution to the paradox is that the information
about the quantum state of collapsing matter be encoded in the
outgoing Hawking radiation \cite{Page},\cite{tHooft}.
This requires a kinematic description of matter at high energies
which differs radically from the one offered by conventional
local quantum field theory.  In the reference frame of a distant
observer the infalling matter
must give up all information about its quantum state to the
emitted Hawking radiation as it approaches the event horizon.
There is, however, no invariant local signature of the presence
of an event horizon: the curvature remains smooth there, and an
observer in free fall who enters a large black hole will not
experience any discomfort upon crossing the horizon.
The principle of black hole complementarity \cite{STU} states that
these seemingly contradictory viewpoints are equally valid.
Reference \cite{STU} put forward a phenomenological description of
black hole evolution in terms of a quantum mechanical membrane,
the ``stretched horizon'', which absorbs the energy and quantum
information of incoming matter, thermalizes it, and eventually
radiates it back as Hawking radiation.

These ideas are at odds with
the usual semi-classical approach to black hole physics,
where one considers quantum evolution of local fields in a
(semi-)classical background geometry.  Analysis of several
gedanken experiments, designed to test
black hole complementarity, indicates, however, that it does not
contradict any known laws of physics, and that the conflict
can be traced to unwarranted assumptions about short-distance
physics which are implicitly made in the usual semi-classical
approach \cite{Gedanken}.

String theory is widely believed to provide a consistent
short-distance description of matter and gravity and Susskind
has recently argued that relativistic strings in fact exhibit
precisely the kinematic behavior required to implement black hole
complementarity \cite{Susskind}.  Following this work, we study
relativistic strings approaching the event horizon of a black hole.
We show that the transverse spreading of the string configuration
simulates a stochastic process.  More specifically, we obtain a
Langevin equation describing the coarse-grained transverse
evolution of the string and interpret it in terms of a branching
diffusion of string bits.  The infalling matter is thus very
efficiently thermalized as it is absorbed into the stretched
horizon.

Let us begin by briefly reviewing the arguments for the
spreading of strings falling into a black hole.  The key
observation is that, due to zero-point fluctuations of string
modes, the size and shape of strings are sensitive to the time
resolution used \cite{Susskind2},\cite{KKS}.  The shorter the
time over which the oscillations of a string are averaged
the larger is its spatial extent.

Consider a string configuration in free fall approaching a
black hole event horizon.  An observer at rest far away from
the black hole measures asymptotic time, but because of the
increasing redshift, a unit of asymptotic time corresponds to
an ever shorter time interval in the free-fall frame.  The
distant observer is therefore using a shorter and
shorter resolution time to describe the string configuration
and, once it passes within a proper distance of order the
string scale from the event horizon, the string begins to
spread both in the longitudinal and transverse directions.
The longitudinal spread is sufficiently rapid to
cancel out the longitudinal Lorentz contraction caused by the
black hole geometry.
Meanwhile, the spread in the transverse directions causes the
configuration to cover the entire horizon area in a time
which is short compared to the black hole lifetime.
The stretched horizon is thus {\it made out of\/} the strings
in the infalling matter which forms the black hole.

On the other hand, this spreading effect is not present
in the free-fall frame, where there is no redshift
to enhance the time resolution, and from the point of view
of an infalling observer there is no stretched horizon, in line
with the principle of black hole complementarity.

A natural question to ask is in what sense does an asymptotic
observer ``see'' the string spreading.  First of all, any
direct observation involving the scattering of test strings
off the infalling matter near the event horizon, which then
propagate to infinity, requires an enormous center of mass
energy due to the gravitational redshift.  The results of
such scattering experiments would therefore be consistent
with the notion that the infalling string has spread out.
In the absence of such high energy scattering experiments,
distant observers can only make measurements on the outgoing
Hawking radiation, and the question is whether measurements
will uncover subtle correlations amongst the outgoing
quanta.  The mathematical description of the string
spreading suggests the existence of physical processes
capable of producing such correlations.

Since the stringy stretched horizon is formed from the infalling
matter itself, it efficiently absorbs the quantum information
contained in that matter.  Our results below illustrate how the
string spreading process also thermalizes the stretched horizon.
This comes about because a time-dependent cutoff is being used
on the scalar fields which give the transverse location of the
string.  As time goes on, new modes emerge below the cutoff,
and the random phases of the different modes lead to
a {\it classical} stochastic evolution of the transverse fields.
The physics of the process is quite analogous to that
of field evolution in the rapidly expanding background geometry
of an inflationary universe; and we shall draw upon the cosmology
literature in our analysis.
For technical reasons our discussion is limited to strings
approaching the horizon of a classical black hole in the
limit of large mass.  Therefore we cannot at present establish,
although we find it very plausible, that the stringy stretched
horizon re-emits the original information encrypted in apparently
thermal Hawking radiation.

\section{Infalling string near the horizon}

The string spreading effects in which we are interested take
place on a short timescale compared to the black hole lifetime
\cite{Susskind} and we therefore consider a static classical
geometry.  Furthermore, the spreading takes place in a thin
layer (whose proper thickness is of order the string scale
$\sim {1\over gM_{Pl}}$) outside the event horizon and
for a macroscopic black hole this region is well approximated
by Rindler space.  The Schwarzschild line element can be
written\footnote{Throughout the paper we will use units
where $M_{Pl}=1$.}
\begin{equation}\label{linel}
ds^2 = -{2M\over r}e^{-(r/2M -1)}\>dU\,dV
+ r^2 \,d\Omega^2 \>,
\end{equation}
where the Kruskal variables $U,V$ are defined in terms of the
original Schwarzschild coordinates through
\begin{equation}
-{U\over V} = e^{t/2M}\>,\qquad
-U\,V = 16M^2\,({r\over 2M}-1)\,e^{(r/2M -1)} \>.
\end{equation}
As the horizon is approached its spherical shape is well
approximated locally by a planar surface, so near $r=2M$ we
can write,
\begin{equation}
ds^2 = - dU\,dV + d{X^\perp}^2 \>.
\end{equation}
The null coordinates $U,V$ extend into the black hole interior
and are appropriate for an observer passing through the horizon
in free fall.  They are related to the null coordinates $u,v$ of
an asymptotic observer through,
\begin{equation}
{U\over 4M}=-e^{-u/4M}\>,\qquad
{V\over 4M}=e^{v/4M} \>.
\end{equation}
The usual description of Rindler space is obtained by rescaling
all the coordinates to absorb the factors of $4M$, but we prefer
to keep these factors explicit.

Rindler space is isomorphic to a slice of flat Minkowski space,
where we can easily discuss free string propagation in light-cone
gauge.  Consider an infalling string described by transverse
coordinates $X^i(\sigma,\tau)$, with $i=1,2$, and some
internal degrees of
freedom depending on the string model being used.
In the free-fall frame the light-cone gauge condition is
$\tau=U/4M$.  The internal degrees of freedom decouple from the
transverse coordinates which satisfy a free wave equation,
\begin{equation}\label{waveeq}
\Biggl[   \frac{\partial^2}{\partial \tau^2}
-\frac{\partial^2}{\partial \sigma^2} \Biggr]
X^i (\sigma,\tau) = 0 \>.
\end{equation}
The solution can be expressed by the usual sum over modes of
oscillation, but, unless the infinite sum is cut off in some way,
this leads to ill-defined expressions for quantities such as the
average transverse area occupied by the string in the quantum
theory \cite{Susskind2},\cite{KKS}.
Introducing a cutoff on the mode expansion corresponds
physically to employing finite time resolution so that mode
oscillations above a given frequency average out.
The string wave-function extends over a transverse area which
grows logarithmically with better resolution time while the
average length of string projected onto the transverse
directions grows linearly \cite{KKS}.  This means that the string
density will increase at the center of the distribution, and
eventually we can no longer neglect the effect of string
interactions there.  Our calculations, which are at the level of
free string theory, will give a good description of the spreading
process at the center early on and, as it turns out, remain
valid at the outer edges of the spreading string configuration.

Now we want to consider the string evolution in the reference
frame of a distant fiducial observer, whose retarded time,
$t=u$, is related to the worldsheet parameter time through
\begin{equation}\label{timerel}
{\tau} = - e^{-{t / 4M}} \>.
\end{equation}
As asymptotic time goes on, a fixed resolution in $t$ thus
corresponds to an exponentially improving resolution in $\tau$,
so that in the reference frame of the distant observer, the string
wave-function will spread with time to occupy a transverse
area proportional to $t$.
By using the constraint equations of light-cone string theory
one can show that the string also spreads in the longitudinal
direction and that the longitudinal spreading is rapid enough
to balance the Lorentz contraction due to the black hole
metric near the event horizon \cite{Susskind}.

Let us look at the transverse evolution of the string in more
detail.  The following discussion closely parallels the treatment
of scalar perturbations in models of inflationary cosmology
(see for example \cite{Star},\cite{Sasaki},\cite{SJRey}).
In asymptotic time the equation of motion (\ref{waveeq}) of
the field $X^i\bigl(\sigma,\tau(t)\bigr)$ becomes
\begin{equation}\label{tdeforx}
\Biggl[   \frac{\partial^2}{\partial t^2}+
\frac{1}{4 M}\frac{\partial}{\partial t} -
\left(\frac{e^{-t/4M}}{4M} \right)^2
\frac{\partial^2}{\partial \sigma^2} \Biggr]
X^i (\sigma,t) = 0 \>.
\end{equation}

Acting along the lines of \cite{Sasaki}, we split both
the field
$X^i$ and its conjugate momentum
$\dot{X}^i$ $( \hskip5pt \dot{} \equiv \frac{\partial}{\partial t})$
into a
slowly-varying, classical part and fast-varying, quantum part
\begin{eqnarray}
X^i (\sigma,t) &=& x^i (\sigma,t) + x_f^i (\sigma,t) \,, \nonumber\\
\dot{X }^i (\sigma,t) &=& v^i (\sigma,t) + v_f^i (\sigma,t) \, .
\end{eqnarray}
The quantum field $x_f^i$ can be expressed as a sum over modes in
the $(\tau,\sigma)$ frame, provided the frequency cutoff, which
separates the fast modes from the slow ones, is chosen to reflect
the exponentially improving resolution in $\tau$:
\footnote{Our conventions are
$[c_n^i,c_m^{j \dagger}]=\delta_{nm}\delta^{ij}
= [\tilde{c}_n^i,\tilde{c}_m^{j \dagger}]$,
all other commutators zero.}
\begin{equation}\label{qmodeexp}
x_f^i (\sigma,t) = \sum_{n=1}^\infty W(n+\frac{\epsilon}{\tau})
\Biggl[ \frac{c_n^i}{\sqrt{n}} \, x_n^+ +
\frac{\tilde{c}_n^i}{\sqrt{n}} \, x_n^- + h.c. \Biggr] \>,
\end{equation}
where we have defined $x_n^{\pm} = \sqrt{\frac{\alpha'}{2}} \,
e^{-in(\tau \pm \sigma)}$, and $\epsilon$ is some constant.
For a filter function we could use
$W(n+\frac{\epsilon}{\tau})=\theta(n+\frac{\epsilon}{\tau})
=\theta(n-\epsilon \, e^{t/4M})$; then an asymptotic observer would
indeed include only modes of exponentially higher frequency in the
definition of the quantum part of the field.  In our
calculations we will use a filter where the step of the theta
function is smeared a little, in order to avoid unphysical effects
associated with a sharp edge cutoff in two dimensions.

The expansion for $v_f^i$ is similar:
\begin{equation}
v_f^i (\sigma,t) = \sum_{n=1}^\infty W(n+\frac{\epsilon}{\tau})
\Biggl[ \frac{c_n^i}{\sqrt{n}} \, \dot{x}_n^+
+ \frac{\tilde{c}_n^i}{\sqrt{n}} \, \dot{x}_n^- + h.c. \Biggr] \, .
\end{equation}

With the above definitions, the long-wavelength fields, $x^i$ and
$v^i$, will evolve nontrivially in asymptotic time due to the
continual feeding in of modes from the quantum parts.
In order to obtain an equation for this evolution, we
substitute $X^i = x^i + x_f^i$ and $\dot{X}^i = v^i + v_f^i$
into the field equation (\ref{tdeforx}).
First, let us write
\begin{equation}
\dot{x}_f^i = \sum_{n>0} W(n+\frac{\epsilon}{\tau})
\Biggl[ \frac{c_n^i}{\sqrt{n}} \, \dot{x}_n^+
+ \frac{\tilde{c}_n^i}{\sqrt{n}} \, \dot{x}_n^- + h.c. \Biggr] -
\eta^i(\sigma,t)
\end{equation}
where $\eta^i(\sigma,t)$ is defined as
\begin{equation}\label {etadefn}
\eta^i(\sigma,t) \equiv - \frac{1}{4M} \sum_{n>0}
\frac{\epsilon}{\tau} \, W^{'} (n+\frac{\epsilon}{\tau})
\Biggl[ \frac{c_n^i}{\sqrt{n}} \, x_n^+ +
\frac{\tilde{c}_n^i}{\sqrt{n}} \, x_n^- + h.c. \Biggr] \, ,
\end{equation}
and similarly,
\begin{equation}
\dot{v}_f^i = \sum_{n>0} W(n+\frac{\epsilon}{\tau})
\Biggl[ \frac{c_n^i}{\sqrt{n}} \, \ddot{x}_n^+
+ \frac{\tilde{c}_n^i}{\sqrt{n}} \, \ddot{x}_n^- + h.c. \Biggr] -
\xi^i(\sigma,t)
\end{equation}
where $\xi^i(\sigma,t)$ is defined as
\begin{equation}\label{xidefn}
\xi^i(\sigma,t) \equiv - \frac{1}{4M} \sum_{n>0}
\frac{\epsilon}{\tau} \,
W^{'} (n+\frac{\epsilon}{\tau}) \Biggl[ \frac{c_n^i}{\sqrt{n}}
\, \dot{x}_n^+ + \frac{\tilde{c}_n^i}{\sqrt{n}} \, \dot{x}_n^- + h.c.
\Biggr]  \, .
\end{equation}
Upon making these substitutions, the field equation for $X^i$
reduces to two coupled equations for the long-wavelength fields:
\begin{eqnarray}\label{xveqns}
\dot{x}^i &=& v^i + \eta^i \> , \nonumber\\
\dot{v}^i &=& - \frac{1}{4M} v^i
 + \frac{\tau^2}{(4M)^2}
 \frac{\partial^2}{\partial \sigma^2} x^i + \xi^i \, .
\end{eqnarray}
Near the horizon $\tau\rightarrow 0$ and so the spatial
derivative term becomes negligible.  In order to know which of
the other terms are important, we must study the quantum
noise functions $\eta$ and $\xi$ as defined in eq.(\ref{etadefn})
and eq.(\ref{xidefn}).

Let us first consider what to use for the filter function.
Were we to use a simple step-function,
$W=\theta(n+\frac{\epsilon}{\tau})$,
we would have
$- \frac{\epsilon}{\tau}  \, W^{'} (n+\frac{\epsilon}{\tau})=
\delta (n\frac{\tau}{\epsilon} +1)$.
Instead, we use a smooth approximation to this $\delta$-function,
a gaussian function of $n$ centred about
$\frac{\epsilon}{| \tau |}$ and with width
$\frac{\beta \epsilon}{| \tau |}$ where $\beta$
is a small number:
\begin{equation}
-\frac{\epsilon}{\tau} \, W^{'} (n+\frac{\epsilon}{\tau},\beta) =
\frac{1}{\sqrt{2\pi} \beta}
\exp \biggl[-\frac{1}{2 \beta^2} (\frac{n \tau}{\epsilon}+1)^2
\biggr] \ .
\end{equation}
Given a filter function of this form we can compute the various
noise correlation functions and commutators.  We first obtain the
$\eta$ correlator:
\begin{eqnarray}\label{etacorr}
 \langle \eta^i(1) \,\eta^j(2) \rangle &=&
\frac{1}{(4M)^2} \sum_{m,n>0}
\frac{1}{2\pi\beta^2}
\exp \biggl[-\frac{1}{2 \beta^2} (\frac{m \tau_1}{\epsilon}+1)^2
-\frac{1}{2 \beta^2} (\frac{n \tau_2}{\epsilon}+1)^2 \biggr]
\times
\nonumber\\
&\times& \langle 0 | \Biggl[
\frac{c^i_n}{\sqrt{n}} \, x_n^+(1) + \frac{\tilde{c}^i_n}{\sqrt{n}}
\, x_n^-(1) + h.c. \Biggr]
\Biggl[ \frac{c^j_m}{\sqrt{m}} \, x_m^+(2) +
\frac{\tilde{c}^j_m}{\sqrt{m}}
\, x_m^-(2) + h.c.
\Biggr] | 0 \rangle \>. \nonumber\\
& &
\end{eqnarray}
The expectation value is
\begin{equation}
\langle 0 | \cdots | 0 \rangle
=  \frac{\alpha'}{n} \delta^{ij}\delta_{mn}
e^{-in(\tau_1-\tau_2)}
\,  \cos[n(\sigma_1-\sigma_2)] \>,
\end{equation}
so that the double sum in eq.(\ref{etacorr}) reduces to a single sum.
The identity
\begin{equation}
(\frac{n \tau_1}{\epsilon}+1)^2 + (\frac{n \tau_2}{\epsilon}+1)^2
=2 (\frac{n \bar\tau}{\epsilon}+1)^2
+ \frac{n^2}{2\epsilon^2}
(\tau_1-\tau_2)^2  \>,
\end{equation}
where $\bar\tau = \frac{1}{2}(\tau_1+\tau_2)$,
allows us to rewrite the correlation function as
\begin{eqnarray}\label{etas}
 \langle \eta^i(1) \,\eta^j(2) \rangle &=&
\frac{\alpha'}{2}\, \frac{\delta^{ij}}{(4M)^2} \sum_{n>0}
\frac{1}{\pi\beta^2 n}
\exp \biggl[-\frac{1}{\beta^2}
(\frac{n \bar\tau}{\epsilon}+1)^2 \biggr]
\exp \biggl[-\frac{n^2}{4\beta^2\epsilon^2}
(\tau_1-\tau_2)^2 \biggr]
\times
\nonumber\\
&&\ \times\
e^{-i n (\tau_1-\tau_2)} \cos[n(\sigma_1-\sigma_2)] \>.
\end{eqnarray}
One of the gaussian factors gives a $\delta$-function in time
for small $\beta$.  Near the horizon, where
$\bar\tau\rightarrow 0$, the remaining sum is
well approximated by an integral and we obtain:
\begin{equation}
\label{noisecorr}
\langle \eta^i(1) \, \eta^j(2) \rangle \simeq
\frac{\alpha'}{2}\, \frac{\delta^{ij}}{4 M} \,
\delta(t_1-t_2) \, \,  \cos(\frac{\epsilon \Delta \sigma}{\tau}) \,
\exp \biggl[-\frac{\beta^2}{4} (\frac{\epsilon \Delta
\sigma}{\tau})^2 \biggr] \>.
\end{equation}
Had we used a sharp step-function as our filter function,
{\it i.e.} taken $\beta=0$, we would have obtained a purely
oscillatory
spatial dependence for the noise correlator,
an unphysical artefact of using an abrupt cutoff.

Computing the $\xi$ correlator involves the same steps and
we find that
\begin{equation}\label{xixi}
\langle \xi^i(1) \, \xi^j(2) \rangle \simeq
-\Bigl(\frac{\epsilon}{4M}\Bigr)^2\,
\langle \eta^i(1) \, \eta^j(2) \rangle \>.
\end{equation}
We can also calculate the various commutators of $\eta$ and $\xi$
at the same level of approximation, with the results
$[\eta^i(1),\eta^j(2)] = 0 = [\xi^i(1),\xi^j(2)]$ and
\begin{equation}\label{xietacomm}
[\eta^i(1),\xi^j(2)] = -\frac{i\epsilon}{4M} \,
\langle \eta^i(1) \, \eta^j(2) \rangle \, .
\end{equation}
For a macroscopic black hole $M \gg \epsilon$, so from
(\ref{xixi}) and (\ref{xietacomm})
we see that only the $\eta^i$ noise term is important in
eq.(\ref{xveqns}).  The fact that the commutators
of the fields are small as compared with their correlators
shows that $x^i$ and $v^i$ become effectively classical.

The noise $\xi^i$ in the momentum equation is negligible in the
$\epsilon \ll M$ limit and can be dropped.  The equation for $v^i$
is then solved by $v^i=-\tau v^i_0$, which means that near the
horizon where $\tau\rightarrow 0$ we may ignore $v^i$ altogether.

Thus the coupled equations (\ref{xveqns}) reduce to a single,
classical, Langevin equation for the transverse position
$x^i$,
with quantum noise $\eta^i$:
\begin{equation}\label{Langevin}
\dot{x}^i =  \eta^i \>.
\end{equation}
The time-dependence in the correlator (\ref{noisecorr}) assures us
that the noise $\eta^i$ is white, and the spatial dependence tells us
that wee bits of string of parameter length $\Delta \sigma \simeq
| \tau | / (\beta \epsilon)$ evolve independently with time.

\section{Branching diffusion of string bits}

An important insight from the physics of the inflationary universe
is that the Langevin equation of the previous section can be
given an interpretation in term of a {\it branching} diffusion
process \cite{LinMezh}.  Indeed,  for any given point of the string,
equations (\ref{noisecorr}) and (\ref{Langevin}) tell us that the
value of the slowly-varying field $x^i(\sigma, t)$ experiences a
Brownian motion which is essentially unaffected by anything
lying outside the correlation length
$\Delta \sigma \approx \frac{|\tau |}{\beta \epsilon}$\@. This
resembles the independence of the scalar field evolution in different
Hubble domains in the case of chaotic inflation.  The correlation
length decreases exponentially with asymptotic time and therefore
the number of such independent bits of the string increases
exponentially:
\begin{equation}\label{number}
N \sim \frac{2\pi\beta\epsilon}{| \tau |} =
2\pi \beta\epsilon \, \exp(t/4M) \, .
\end{equation}
We can describe the process as a branching diffusion where every bit
of the string diffuses independently of all others and splits into
two\footnote{This particular ratio of 2:1 of the number of
``daughter'' bits per ``parent'' corresponds to a specific time
interval of effective temporal averaging. However, as in the case of
chaotic inflation \cite{LinMezh}, this parameter drops out from the
equations describing the average number of string bits.} bits with
intensity $\rho(x)$, which is determined by matching the growth of
the number of bits in eq. (\ref{number}).
The two bits then diffuse away from their ``birthplace'', independent
of each other, and split in their turn, and so on.

The diffusion of a given bit of string is governed by the
Fokker-Planck equation:
\begin{equation}
\frac{\partial}{\partial t} P(x^i,t) = \frac{\varsigma^2}{2}
\delta^{ij}
\frac{\partial}{\partial x^i}
\frac{\partial}{\partial x^j} P(x^i,t) \, .
\end{equation}
where $P(x^i,t)$ is the normalized probability of finding that
bit at $(x^i,t)$.  On the stretched horizon of a finite-mass
black hole, $P \rightarrow const$ at late times.

The coefficient of diffusion can be read off the correlator of the
noise, $\varsigma ^2 = \frac{\alpha'}{8M}$, and so the effective
temperature of this diffusion is, according to the asymptotic
observer, given by $T \sim \frac{1}{M}$.

This branching diffusion picture has implications for various
properties of the infalling string and  these can be checked
against previous results \cite{Susskind},\cite{KKS}.
Consider first the mean square transverse position,
\begin{equation}\label{xsqt}
\langle x^i x^i \rangle = 2 \varsigma^2 t
=\frac{\alpha'}{4M} \,  t \, .
\end{equation}
This is the transverse spread, linear in asymptotic time, pointed
out by Susskind \cite{Susskind}.

The equation for the average number of branching bits
$\cal N$ at a time $t$ is the Kolmogorov-Fokker-Planck equation
(see \cite{LinMezh} for a derivation in the case of chaotic
inflation), which in our case turns out to be simply
\begin{equation}
\frac{\partial}{\partial t} {\cal N} = \frac{\varsigma^2}{2}
\delta^{ij}
\frac{\partial}{\partial x^i} \frac{\partial}{\partial x^j} {\cal N}
+ \rho(x) {\cal N} \, .
\end{equation}
Choosing the branching intensity to be the constant
 $\rho = 1/4M$, we get, asymptotically,
\begin{equation}\label{avgN}
{\cal N} \sim \exp(\frac{t}{4M}) \, .
\end{equation}
which does indeed match the behavior (\ref{number}).

This gives us the average time between splitting as
\begin{equation}
t_{spl} \sim \frac{1}{\rho} = 4M \, .
\end{equation}
Therefore, using (\ref{xsqt}), we see that the average transverse
distance between two adjacent string bits is
\begin{equation}
l_{*} \sim \sqrt{\frac{\alpha'}{4M} \,  t_{spl}}
\sim \sqrt{\alpha'} \> ,
\end{equation}
which finally gives us the average transverse length of the string:
\begin{equation}
{\cal L} = l_{*} \, {\cal N}
\sim \sqrt{\alpha'}  \,
\exp(\frac{t}{4M}) \, .
\end{equation}

These considerations can be extended in a straightforward manner to
the branching diffusion process which describes string spreading
across the horizon of a higher dimensional analog of a Schwarzschild
black hole.  In $D$ dimensions, the Hawking temperature in Planck
units of such a black hole is given by
$2\pi T_D = f(D) \, (\frac{1}{M})^\frac{1}{D-3}$
where $f(D)$ depends only on $D$ \cite{devegasanchez}.  Thus, we
simply make the replacement $\frac{1}{4M} \rightarrow 2\pi T_D$
in expressions like (\ref{avgN}).
In eq.(\ref{xsqt}) we must also replace $2 \rightarrow (D-2)$,
since we are summing over $i$.  The average transverse distance
between adjacent string bits then becomes:
\begin{equation}\label{length}
l_{*} \sim \sqrt{\alpha'(D-2)} \>,
\end{equation}
where $D$ is the number of extended spacetime dimensions.

We can also estimate the average extrinsic curvature per unit length
of the string using the branching diffusion picture.
Consider at time $t_0$ a piece of a string; we may think of it as
composed of wee bits of string separated by typical distance $l_{*}$.
After a time $\Delta t \sim t_{spl}$ each bit will have
branched into two, with the new bits again being separated by a
length $l_*$ of string.  In this discretized picture the string
looks like a zigzag, see Fig.1(a).

\vskip2.5cm
{\centerline{\epsfsize=3.0in \hskip 0cm \epsfbox{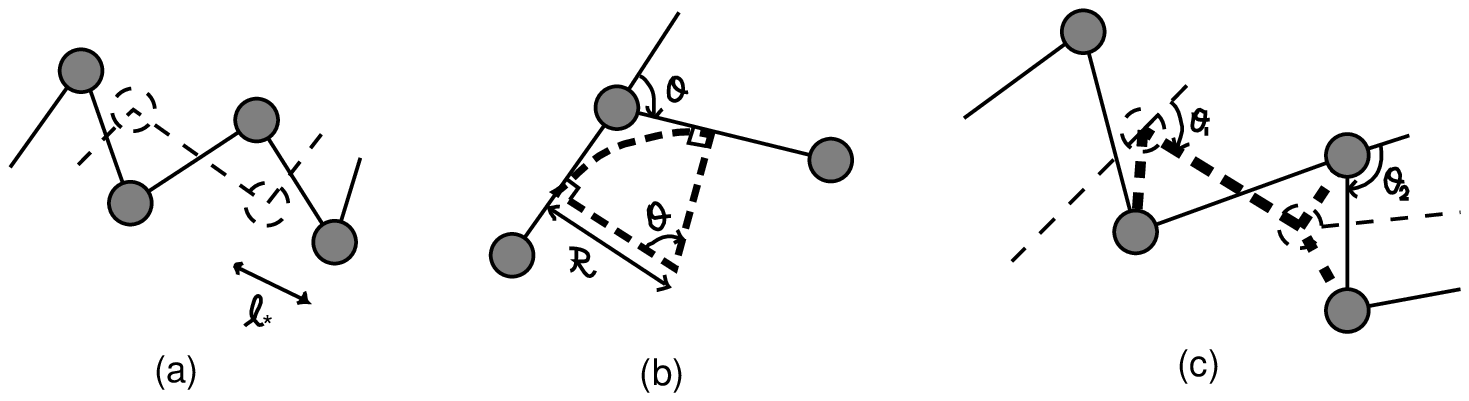}}}
\vskip0cm
{\tenrm{
FIGURE 1. \hfil\break
(a) A piece of string undergoing a branching process.
At first, there are two bits, denoted by the dotted circles; then a
branching takes place and there are now four bits, denoted by the
colored circles.  \hfil\break
(b) The angle $\theta$ indicates how bent the zigzag is.  The
curve of the dotted line is the smoothed zigzag and $\cal R$ is its
radius of curvature.\hfil\break
(c) The four vectors needed to follow a branching process are
indicated by the thick broken lines; other lines are as in (a).
The angles before and after branching are indicated by $\theta_1$
and $\theta_2$.}}

\vskip2.0cm

We would now like to find the analogue of the average curvature along
the string.  Imagine smoothing out each corner of a zigzag; this is
shown in Figure 1(b).  Then the average extrinsic curvature for a
piece of zigzag is
\begin{equation}
\langle R \, \rangle = \int d {l} \,  \frac{1}{\cal R}
= \int {\cal R}  \, d\phi  \, \frac{1}{\cal R} = \theta \, .
\end{equation}
In this picture, then, we are interested in the average value of the
angle $\theta$.  However, to follow the evolution of $\theta$ during
a branching process, we need only four vectors, as can be seen from
Figure 1(c).  Thus, the average angle can depend only on a
four-dimensional subspace of the $(D{-}2)$-dimensional transverse
space.
Therefore, for $D\geq 6$ the average angle
$\theta_*$ is constant as a function of $D$, and the extrinsic
curvature per unit length is
\begin{equation}\label{extrcurv}
r = \frac{\langle R \,  \rangle}{l_*} \sim \frac{\theta_*}{l_*} \sim
\frac{1}{\sqrt{\alpha' (D-2)}} \, .
\end{equation}
The leading $D$ dependence of both the average length in
(\ref{length})
and the extrinsic curvature in (\ref{extrcurv}) agree with what was
found by Karliner {\it et al.} \cite{KKS}.  We have emphasized the
application to black hole physics but branching diffusion evidently
provides a useful physical picture in its own right of how the size
and shape of strings develop as a function of mode cutoff.

\section{Discussion}

In four spacetime dimensions the string will spread to cover the area
of the black hole horizon in a time
\begin{equation}\label{tspread}
t_S\sim g^2 M^3 \>,
\end{equation}
where $M$ and $t$ are measured in Planck units and $g$ is the string
coupling strength \cite{Susskind}.  This is a short time compared to
the black hole lifetime if the string is weakly coupled.\footnote{If
$D>4$ the lifetime $t_L \sim M^\frac{D-1}{D-3}$ is always long
compared to $t_S \sim g^2 M^\frac{3}{D-3}$.}  In the branching
diffusion picture the relation (\ref{tspread}) easily follows from
eq. (\ref{xsqt}).

Another important timescale is that on which the volume density of
string at the centre of the distribution becomes $O(1/g^2)$ in
string units.  At this point string interactions can no longer be
ignored and our calculations, which are all at the level of free
string theory, no longer apply.  The proper thickness of the
stretched horizon remains of order one in string units while the
average area occupied increases linearly with time.  At the same
time the average number of string bits grows exponentially so the
time $t_I$ at which string interactions become important at the
center of the distribution of string bits is:
\begin{equation}\label{tint}
t_I \sim M {\ln}{(\frac{1}{g^2})} \>.
\end{equation}
For a macroscopic black hole this timescale is very much shorter
than the spreading time $t_S$.  An attractive possibility, pointed
out in \cite{Susskind}, is that string interactions prevent the
volume density from exceeding $O(1/g^2)$ and that the central
density will level off at that value.  This volume density
corresponds to an $O(1)$ area density of string bits on the stretched
horizon measured in Planck units, the value suggested by the
Bekenstein-Hawking entropy.  At any rate, the high string density
in the center does not affect the branching diffusion at the fringe
of the distribution so our estimate (\ref{xsqt}) of the transverse
spreading rate remains unchanged.

The aim of the present paper was to provide some additional support
for the view that a stretched horizon made out of strings has a
key role to play in resolving the black hole information paradox.
There remain many open questions within this approach; for example,
how to extend the calculations from Rindler space to finite mass
black hole geometries.  Another important issue to address is the
underlying causal properties of string theory which allow the
spread of information across the horizon, see {\it e.g.} the
forthcoming work of Lowe {\it et al.} \cite{LSU}.

\vskip3cm
\section*{Acknowledgements}
The authors wish to thank D. Lowe, A. Linde, A. Starobinsky
and especially L. Susskind for helpful discussions.
This work was supported in part by National Science Foundation
grants PHY-89-17438 and PHY-89-04035.

\end{document}